\documentclass[showpacs,aps,prl,twocolumn,nofootinbib]{revtex4}
\usepackage{amsmath}
\usepackage{epsfig}
\usepackage{subfigure}
\usepackage{float}
\usepackage{verbatim} 
\usepackage{axodraw4j} 
\usepackage{pstricks}
\usepackage{color}
\usepackage{slashed}
\usepackage{multirow}
\usepackage{pstricks}
\usepackage{color}
\usepackage{booktabs}
\usepackage{subfigure}
\usepackage{epstopdf}

\newcommand{\lag}{{\cal L}}

\newcommand{\hc}{\ensuremath{\mathrm{H.c.}}}
\renewcommand{\lag}[1]{\ensuremath{\mathcal{L}_\mathrm{#1}}}

\newcommand{\DET}{\Gamma(1+\epsilon)\left(\frac{4\pi\mu^2}{m_t^2}\right)^\epsilon}
\newcommand{\DE}{\Gamma(1+\epsilon)\left(4\pi\right)^\epsilon}

\begin{document}

\leftline{}
\rightline{CP3-13-26}
\title{ Top-quark decay into Higgs boson and a light quark at next-to-leading order in QCD}
\author{Cen Zhang, Fabio Maltoni}
\affiliation{
Centre for Cosmology, Particle Physics and Phenomenology, 
Universit\'e Catholique de Louvain, B-1348 Louvain-la-Neuve, Belgium 
}

\begin{abstract}
Neutral flavor-changing transitions are hugely suppressed in the Standard Model
and therefore they are very sensitive to new physics. We consider the decay
rate of $t \to  u_i \, h $ where $u_i=u,c$ using an effective field theory
approach.  We perform the calculation at next-to-leading order (NLO) in QCD
including the relevant dimension-six operators. We find that at NLO the
contribution from the flavor-changing chromomagnetic operator is as important
as the standard QCD correction to the flavor-changing Yukawa coupling.  In
addition to improving the accuracy of the theoretical predictions, the NLO
calculation provides information on the operator mixing under the
renormalization group.
\end{abstract} 


\maketitle

\section{Introduction}

The discovery of a particle of about 125 GeV mass
~\cite{Chatrchyan:2012ufa,Aad:2012tfa} that resembles the Higgs boson of the
Standard Model (SM)~\cite{Englert:1964et,Higgs:1964pj,Weinberg:1975gm} has
opened a new era in particle physics. A new realm of possibilities for
exploring the electroweak breaking sector and new exciting opportunities to
search for new physics in general have appeared. From the existence of new
symmetries to new space-time dimensions, from new matter to a richer scalar
sector, many are still viable options. For instance, extra scalar states could
exist mostly (or exclusively) coupling to the Higgs boson or new particles
could decay into final states involving the SM scalar boson. In addition to
direct searches, the accurate measurement of the coupling strengths and
structures to the SM particles could point to the scale of new physics.

In this perspective, the study of neutral-flavor-changing (NFC) couplings
involving the top quark and the Higgs boson is of special interest. In the
Standard Model, NFC interactions are absent at tree level and hugely suppressed
by the Glashow-Iliopoulos-Maiani mechanism at one loop. Finding evidence for
such processes taking place at measurable rates would basically always imply
new physics not too far from the TeV scales. The recently observed excess of
$B\to D^{(*)}\tau\nu$ \cite{Lees:2012xj} could hint to NFC mediated by the
Higgs boson~\cite{Crivellin:2012ye}.

In this work we consider the decay of a top quark into a light $u_i$ (up or
charm) quark and the Higgs boson, assuming new physics residing at a scale
$\Lambda > m_t$. The SM contribution to the branching ratio is extremely small,
at order
$10^{-13}$--$10^{-15}$~\cite{Eilam:1990zc,Mele:1998ag,AguilarSaavedra:2004wm}.
Indirect bounds on branching ratio $\mathrm{BR}(t\to ch)$ have been set, for
example, in Refs.~\cite{Larios:2004mx,Aranda:2009cd}, and are found to be at
$\sim10^{-3}$ level. Collider searches for these interactions have been
discussed in \cite{AguilarSaavedra:2000aj,AguilarSaavedra:2004wm,
Kao:2011aa,Wang:2012gp,Atwood:2013ica}. The first limit at LHC,
$\mathrm{BR}(t\to ch)<2.7\%$, was given in \cite{Craig:2012vj}.

In this Letter we present the calculation at next-to-leading order (NLO) in QCD
of the inclusive top-quark decay into a Higgs boson via NFC interactions in an
effective field theory (EFT)~\cite{Weinberg:1978kz} approach. We consider all
lowest dimensional operators $O_i$ compatible with the symmetries of the SM,
\begin{flalign} 
  \lag{EFT}=\lag{SM}+\sum_i \frac{C_iO_i}{\Lambda^2}+ \hc \,
\end{flalign} 
where $\Lambda$ represents the scale of new physics. Our calculation
completes the set of NLO results available for 2-body top decays, such as
$t\to bW$ and $t\to cV$ with $V=\gamma, g, Z$
~\cite{Drobnak:2010ej,Drobnak:2010wh,Drobnak:2010by,Zhang:2010bm}, which we
have independently checked. 

\section{SETUP}

A complete and minimal list of dimension-six operators that can be written with
SM fields and are compatible with the symmetries of the SM can be found in
Ref.~\cite{Grzadkowski:2010es}. We use the same operator basis, employing the
following notation for quark fields:
\begin{flalign}
  Q & : \quad \mbox{third-generation left-handed quark doublet}\,,
  \nonumber\\
  q & : \quad \mbox{first- or second-generation left-handed quark doublet}\,,
  \nonumber\\
  t & : \quad \mbox{right-handed top quark}\,,
  \nonumber\\
  u,c & : \quad \mbox{right-handed up and charm quark}\,,
  \nonumber\\
  \varphi & : \quad \mbox{Higgs boson doublet}\,,
  \nonumber
\end{flalign}
and $\tilde{\varphi}=i\sigma^2\varphi$. There are two main Lorentz structures
that contribute to the $t\to u_ih$ decay, the dimension-six Yukawa interaction
$O_{u\varphi}$, and the chromomagnetic operator $O_{uG}$. The latter
contributes only at NLO. Considering the possible flavor assignments, they read
\begin{flalign}
  &O_{uG}^{(1,3)}=y_t g_s(\bar{q}\sigma^{\mu\nu}T^At)\tilde{\varphi}G^A_{\mu\nu}
  \\
  &O_{u\varphi}^{(1,3)}=-y_t^3(\varphi^\dagger\varphi)(\bar{q}t)\tilde\varphi
  \label{eq:Ouf13}
  \\
  &O_{uG}^{(3,1)}=y_t g_s(\bar{Q}\sigma^{\mu\nu}T^Au)\tilde{\varphi}G^A_{\mu\nu}
  \\
  &O_{u\varphi}^{(3,1)}=-y_t^3(\varphi^\dagger\varphi)(\bar{Q}u)\tilde\varphi\,,
\end{flalign}
where superscript $(1,3)$ and $(3,1)$ denote the flavor structure. The
Hermitian conjugates of the $(3,1)$ operators contribute to $t\to u_ih$ with
the opposite chirality of the corresponding $(1,3)$ operators. In addition,
replacing the up-quark field with the charm-quark field gives the same set of
operators with $(2,3)$ and $(3,2)$ flavor structures. 

Note that the operators have been normalized by attaching appropriate factors
of the top-quark Yukawa coupling $y_t$ and the strong coupling $g_s$. The
powers of these factors are determined by requiring that, whenever these
operators give rise to a SM-like vertex, the coupling strength relative to the
SM coupling is always one of the following factor: 
\begin{equation}
  C_i\frac{m_t^2}{\Lambda^2},\quad C_i\frac{m_t   E}{\Lambda^2},\quad
  C_i\frac{E^2}{\Lambda^2},\quad 
\end{equation}
where $E$ is the typical energy of the particles entering the vertex. This
helps to determine the order of the mixing between these operators. With this
convention, the operator mixing induced by a gluon exchange is always of order
$\alpha_s$, even if the gluon vertex comes from an effective operator. In fact
the normalization coefficient of these operators depends on the details of the
full theory beyond $\Lambda$. In short our convention states that for any
bilinear quark operators, we attach a $y_t$ to each Higgs field, and a $g_s$
for each gluon field. We remark that in this work we choose $y_t$ to be
defined in terms of the on-shell top-quark mass $m_t$ 
\begin{equation}
  y_t=\frac{\sqrt{2}m_t}{v}\,. 
\end{equation}
This is just for simplicity. As a result, it does not contribute to the
anomalous dimension of the operators at order $\alpha_s$.

In the following we focus on operators with $(1,3)$ and $(3,1)$ flavor
structure. The extension of the results from $t\to uh$ to $t\to ch$ is trivial.
In addition, since there is no mixing between operators of type $(1,3)$ and
$(3,1)$ we can omit the superscripts $(1,3)$ and $(3,1)$, and only consider the
$(1,3)$ case. Results for $(3,1)$ can be obtained by flipping the chirality of
the quarks.

At the tree level, only $O_{u\varphi}$ contributes. The effective Lagrangian
describing the interaction of a top quark, a light quark and the Higgs boson
$h$ reads
\begin{equation}
  \mathcal{L}_{tuh}=-C_{u\varphi}\frac{m_t^2}{\Lambda^2}
  \frac{3y_t}{\sqrt{2}}(\bar{u}P_Rt)h+\hc \,.
  \label{eq:tuh1}
\end{equation}
In addition, the terms in $O_{u\varphi}$ involving the vacuum expectation value
$v$ of the Higgs field give rise to $u_L-t_R$ mixing. The standard way to deal
with this effect is to perform a set of transformations that diagonalize the
mass matrix:
\begin{flalign}
  &u_L\to u_L+ C_{u\varphi}\frac{m_t^2}{\Lambda^2}t_L \,, \\
  &t_L\to t_L- C_{u\varphi}^*\frac{m_t^2}{\Lambda^2}u_L \,,
\end{flalign}
and similarly for $O_{u\varphi}^{(3,1)}$. As a result Eq.~(\ref{eq:tuh1}) is modified and
the $tuh$ interaction reads
\begin{equation}
  \mathcal{L}'_{tuh}=-C_{u\varphi}\frac{m_t^2}{\Lambda^2}
  \frac{2y_t}{\sqrt{2}}(\bar{u}P_Rt)h+\hc
  \label{eq:tuh2}
\end{equation}
Equivalently, one can add a dimension-four counterterm such that the operator
$O_{u\varphi}$ becomes
\begin{equation}
  O_{u\varphi}\to
  O_{u\varphi}+m_t^2y_t(\bar{q}t)\tilde{\varphi}
  =-y_t^3\left(\varphi^\dagger\varphi-\frac{v^2}{2}\right)(\bar{q}t)\tilde{\varphi}\,,
\end{equation}
and the mixing term disappears.

\begin{figure}[t]
  \begin{center}
  \includegraphics[scale=0.35]{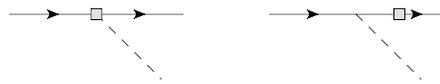}
  \end{center}
  \caption{Feynman diagrams for $t\to u+h$ at tree level. Squares represent an insertion
    of $O_{u\varphi}$.}
  \label{fig:3}
\end{figure}

Another possibility is to keep the quark fields not diagonal, and simply
include the external leg corrections to the diagrams, as shown in
Fig.~\ref{fig:3}. This gives the same result for $t\to uh$. At the one-loop
level, we choose this point of view to take into account the loop-induced $tu$
mixing.

The decay rate up to next-to-leading corrections in the strong coupling can be written as
\begin{equation}
  \Gamma(t\to u_i h) =   \Gamma^{(0)} +   \alpha_s \Gamma^{(1)} \,,
\end{equation}
with LO result \cite{Hou:1991un}
\begin{equation}
  \Gamma^{(0)}=\frac{|C_{u\varphi}|^2}{\Lambda^4}
  \frac{\sqrt{2}G_Fm_t^7}{8\pi}
  \left(1-\frac{m_h^2}{m_t^2}\right)^2 \,,
\end{equation}
where the light quark mass is neglected.

\section{NLO calculation strategy}

We briefly describe our strategy for the computation of NLO corrections in QCD
to the decay rate.

First, we aim at NLO accuracy in QCD but only LO in $C/\Lambda^2$ EFT expansion.
Calculation of higher orders of $C/\Lambda^2$ requires complete knowledge of
dimension-eight operators, and it is beyond the scope of this paper. As there are
no SM FCNC decays $t\to u_ih$ at LO, the first nonzero contribution to the
decay width from new physics is order $(C/\Lambda^2)^2$. 

To regulate both ultraviolet (UV) and infrared (IR) divergences, we employ
dimensional regularization \cite{'tHooft:1972fi} and work in $D=4-2\epsilon$
dimensions. Whenever $\gamma^5$ is present in our computation, we use the
following prescription based on the 't Hooft-Veltman scheme
\cite{Larin:1993tq,Ball:2004rg}: 
\begin{flalign} 
  &\gamma^5\to (1-8a_s)\frac{i}{4!}\epsilon_{\nu_1\nu_2\nu_3\nu_4}
  \gamma^{\nu_1}\gamma^{\nu_2}\gamma^{\nu_3}\gamma^{\nu_4}
  \\
  &\gamma_\mu\gamma^5\to
  (1-4a_s)\frac{i}{3!}\epsilon_{\mu\nu_1\nu_2\nu_3}
  \gamma^{\nu_1}\gamma^{\nu_2}\gamma^{\nu_3}
  \\
  &\sigma_{\mu\nu}\gamma^5\to
  -\frac{i}{2}\epsilon_{\mu\nu\alpha\beta}\sigma^{\alpha\beta}\,.
\end{flalign}
where $a_s=C_F\alpha_s/(4\pi)$.

IR divergences cancel between virtual and real diagrams when sufficiently
inclusive observables are considered. The rest of the calculation involves
the following:
\begin{enumerate}
  \item a UV-divergent part, which gives rise to operator mixing and
    renormalization group equations,
  \item UV-finite part, which gives the actual corrections to the matrix
    elements.
\end{enumerate}

In the first step, we calculate the UV-divergent part arising from the loop
diagrams, and identify the UV counterterms by applying the $\overline{MS}$
scheme and requiring that the UV-divergent terms cancel. The outcome of this
procedure is a set of counterterms for dimension-six operators. We then proceed
to work out the anomalous dimension and the renormalization group equations of
these operators. These equations can be used to evolve the coefficients of
these operators from a higher scale down to the scale of top-quark mass.

In the second step, we calculate the UV-finite part. The final result is
given in terms of the coefficients of these operators defined at the scale of
top-quark mass.

Throughout this paper we ignore the light quark masses, and assume $V_{tb}=1$.

\section{Operator renormalization}

The following counterterms for the SM part are used. For the external fields:
\begin{flalign}
  &\delta Z_2^{(t)}=-\frac{\alpha_s}{3\pi}D_\epsilon\left(\frac{1}{\epsilon_{UV}}
  +\frac{2}{\epsilon_{IR}}+4\right)
  \\
  &\delta Z_2^{(q)}=-\frac{\alpha_s}{3\pi}D_\epsilon\left(\frac{1}{\epsilon_{UV}}
  -\frac{1}{\epsilon_{IR}}\right)
  \\
  &\delta Z_2^{(\varphi)}=0
  \\
  &\delta m_t/m_t=-\frac{\alpha_s}{3\pi}D_\epsilon\left(\frac{3}{\epsilon_{UV}}
  +4\right)
\end{flalign}
while for the couplings:
\begin{flalign}
  &\delta Z_{g_s}=\frac{\alpha_s}{4\pi}\DE\left(\frac{N_f}{3}-\frac{11}{2}\right)
  \frac{1}{\epsilon_{UV}}+\frac{\alpha_s}{12\pi}D_\epsilon\frac{1}{\epsilon_{UV}}
  \\
  &\delta Z_{y_t}=-\frac{\alpha_s}{3\pi}D_\epsilon\left(\frac{3}{\epsilon_{UV}}
  +4\right)
\end{flalign}
where $D_\epsilon\equiv\DET$. This set of counterterms corresponds to
renormalizing the external fields and the top Yukawa coupling on shell, and the
strong coupling in the $\overline{MS}$ scheme. We consider five light flavors
in the running of $\alpha_s$. We then apply the $\overline{MS}$ scheme to the
dimension-six operators and require that dimension-six operators only mix with
dimension-six operators. The counterterms are given by
\begin{equation}
  C^0_i\to Z_{i,j}C_j=(\mathbf{1}+\delta Z)_{i,j}C_j \,.
\end{equation}
We first consider $O_{u\varphi}$. Including counterterms, the Lagrangian can be
written as
\begin{flalign}
  \lag{Eff}=&\mathcal{L}_{tu}+\mathcal{L}_{tuh}
  \\
  \mathcal{L}_{tu}&=-\frac{C_{u\varphi}m_t^3}{\Lambda^2}(\bar{u}_Lt_R)(1+\delta Z_{u\varphi})
  \\
  \mathcal{L}_{tuh}&=-\frac{C_{u\varphi}m_t^2}{\Lambda^2}\frac{3y_t}{\sqrt{2}}
  (\bar{u}_Lt_Rh)(1+\delta Z_{u\varphi})\,,
\end{flalign}
where the counterterm $\delta Z_{u\varphi}$ is
\begin{flalign}
  \delta Z_{u\varphi}&=\delta Z_{u\varphi,u\varphi}
  +\frac{1}{2}Z_2^{(t)}+\frac{1}{2}\delta Z_2^{(q)}
  \nonumber
  \label{eq:deltaZ}\\
  &=\delta Z_{u\varphi,u\varphi}-\frac{\alpha_s}{3\pi}\frac{1}{\epsilon}+\cdots\,.
  \end{flalign}
where the dots stand for additional finite terms. The renormalization mirrors
that of the SM Yukawa terms. $\delta Z_{u\varphi,u\varphi}$ is determined by
the $1/\epsilon$ part of the two-point ($ut$) and three-point ($uth$)
functions. We find
\begin{equation}
  \delta Z_{u\varphi,u\varphi}=-\frac{\alpha_s}{\pi}\frac{1}{\epsilon}\,.
\end{equation}
Now we consider $O_{uG}$. The $utg$ vertex is given by
\begin{flalign}
  \lag{Eff}=-\frac{C_{uG}}{\Lambda^2}2m_tg_s\left(\bar{u}_L\sigma^{\mu\nu}T^At_R\right)
  \partial_\nu G_\mu^A\,. 
\end{flalign}
This gives rise to a $u-t$ mixing at one loop. We find
\begin{flalign}
  \Pi_{ut}(p^2)=-\frac{C_{uG}m_t}{\Lambda^2}\left[
    m_t^2\left(\delta Z_{u\varphi,uG}+\frac{2\alpha_s}{\pi}\frac{1}{\epsilon}
    \right)P_R
    \right.\nonumber\\\left.
    -P_R\left(2p^2-m_t\slashed{p}\right)\frac{a_s}{\pi}\frac{1}{\epsilon}
    \right]
    +\ldots
\end{flalign}
The first term in the bracket implies
\begin{equation}
  \delta Z_{u\varphi,uG}=-\frac{2\alpha_s}{\pi}\frac{1}{\epsilon}\,.
  \label{eq:ZufuG}
\end{equation}
One could also determine this counterterm by calculating the three-point
($uth$) function. The pole in the second term can be dealt through the
following dimension-six counterterms:
\begin{flalign}
  &O_{(1)}=-y_t\left(\bar{q}\overleftarrow{\slashed{D}}
  \slashed{D}t\right)\tilde{\varphi}\,,
  \\
  &O_{(2)}=-\frac{i}{2}y_t^2\left( \bar{Q}\tilde{\varphi} \right)
  \left( \tilde{\varphi}^\dagger \slashed{D}q \right)\,.
\end{flalign}
However, these operators vanish when the equations of motions are considered
(on-shell or off-shell quark does not matter). Therefore one can simply ignore
these operators as the $1/\epsilon$ poles always cancel out when combining with
the vertex contributions in a physical amplitude.

 \begin{figure*}[tb!]
    \begin{center}
      \includegraphics[scale=0.7]{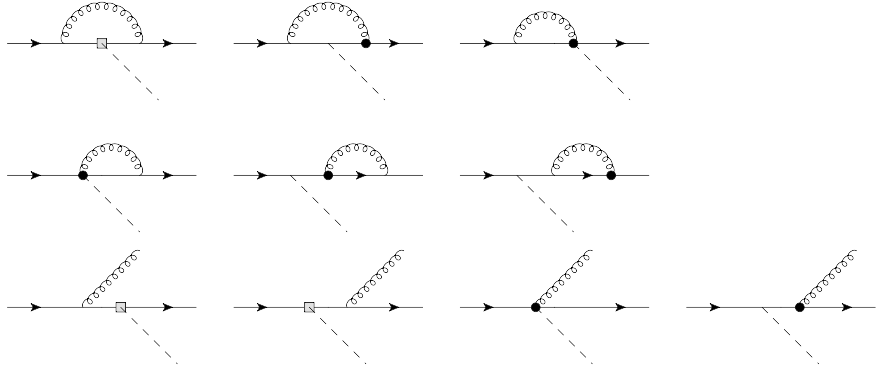}
    \caption{Virtual and real corrections for $t\to uh$. The squares represent the
      contribution from $O_{u\varphi}$, while the black dots represent the contribution
      from $O_{uG}$.}
    \label{fig:4}
    \end{center}
  \end{figure*}

  \begin{figure*}[t]
    \begin{center}
      \includegraphics[scale=0.7]{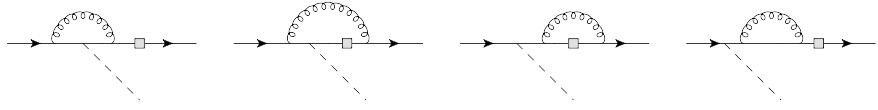}
    \caption{Additional corrections for $t\to uh$.}
    \label{fig:5}
    \end{center}
  \end{figure*}


Finally, the renormalization of $O_{uG}$ requires computation of $utg$ at one
loop. We find:
\begin{equation}
  \delta Z_{uG,uG}=\frac{\alpha_s}{6\pi}\frac{1}{\epsilon}\,,
\end{equation}
while $Z_{uG,u\varphi}$ is zero because there is no contribution from
$O_{u\varphi}$ to $utg$ at one loop at order $\alpha_s$.

In summary, we have the following anomalous dimension matrix for $O_{uG}$ and
$O_{u\varphi}$
\begin{equation}
  \gamma=\frac{2\alpha_s}{\pi}\left(
  \begin{array}{cc}
    \frac{1}{6} &0   \\
    -2          &-1   
  \end{array}\right)\,.
  \label{eq:AD2}
\end{equation}
The operator $O_{uG}$ can also renormalize other operators. The complete
operator mixing can be extracted from the calculation of $t\to u_iV$. However,
at order $\alpha_s$ no other operator can renormalize $O_{u\varphi}$ and
$O_{uG}$, so for the process $t\to u_ih$ the anomalous dimensions given in
Eq.~(\ref{eq:AD2}) are sufficient.

\section{Finite corrections}

We proceed to carry out the UV-finite part of the calculation. At NLO, the loop
corrections and real corrections are shown in Fig.~\ref{fig:4}. To simplify
the calculation, we rotate the $u$ and $t$ quark fields to remove the mixing due
to $O_{u\varphi}$. More specifically, given the following effective Lagrangian,
\begin{flalign}
  \lag{Eff}=&
  \frac{C_{u\varphi}}{\Lambda^2}(-y_t^3)(\varphi^\dagger\varphi)
  (\bar{q}t\tilde{\varphi}) \nonumber  \\
    & \qquad \times \left[ 1+\delta Z_{u\varphi,u\varphi}
    +\frac{1}{2}\delta Z_2^t+\frac{1}{2}\delta Z_2^q\right]
    \nonumber\\
    &+ \frac{C_{uG}}{\Lambda^2}\left[
      y_tg_s(\bar{q}\sigma^{\mu\nu}T^At)G_{\mu\nu}^A \right. \nonumber\\
     &\qquad +\left.\delta Z_{u\varphi,uG}(-y_t^3)(\varphi^\dagger\varphi)(\bar{q}t\tilde{\varphi}
      )\right]
      \label{eq:LEFT2}
\end{flalign}
we perform the following rotation:
\begin{flalign}
  &u_L\to u_L+ C_{u\varphi}\frac{m_t^2}{\Lambda^2}
  \left[ 1+\delta Z_{u\varphi,u\varphi}
    +\frac{1}{2}\delta Z_2^t+\frac{1}{2}\delta Z_2^q\right]t_L \\
  &t_L\to t_L- C_{u\varphi}^*\frac{m_t^2}{\Lambda^2}
  \left[ 1+\delta Z_{u\varphi,u\varphi}
    +\frac{1}{2}\delta Z_2^t+\frac{1}{2}\delta Z_2^q\right]u_L
\end{flalign}
As a result the $u-t$ mixing due to the first term in Eq.~(\ref{eq:LEFT2}) is
removed; therefore, only the vertex correction needs to be considered for
$O_{u\varphi}$. On the other hand, for $O_{uG}$, the second terms in
Eq.~(\ref{eq:LEFT2}) contains a $u-t$ mixing counterterm that is not rotated
away. This term cancels the UV-divergent terms from $O_{uG}$ at one loop in
$u-t$ mixing. The remaining finite part is included in the leg correction
diagrams in Fig.~\ref{fig:4}. Because the rotation of fields is of order
$C/\Lambda^2$, and the LO contribution to $t\to uh$ is already of order
$C/\Lambda^2$, the rotation does not affect our results. As a check, we have
computed the loop corrections without redefining the fields. In this case there
are four more diagrams, as shown in Fig.~\ref{fig:5}.
 
Including both virtual and real corrections, the total NLO correction to the
decay rate is ($x=m_h/m_t$) :
\begin{widetext}
\begin{flalign}
  \frac{8\pi\Gamma^{(1)}}{\sqrt{2}G_Fm_t^7}=&
  \frac{|C_{uG}|^2}{\Lambda^4}\frac{1}{36\pi}
  \Big[ x^8-8x^6-342x^4+620x^2-271
    \nonumber\\& \qquad
    +6x\sqrt{4-x^2}(26-5x^2)\left( \pi-6\sin^{-1}\frac{x}{2} \right)
    +12(9x^4+76x^2-8)\log x\Big] \nonumber
  \\
  &-\frac{\mathrm{Re}(C_{uG}C_{u\varphi}^*)}{\Lambda^4}\frac{2}{9\pi}
\bigg\{
  6\bigg[ 6(1-x^2)^2\log\frac{m_t}{\mu}+(5x^4+2x^2+4\log(1-x^2)-2\log x)\log x
    \nonumber\\& \qquad
    +\left( \sqrt{\frac{4}{x^2}-1}\, (x^4-6x^2+8)+2\pi \right)
    \arcsin\frac{x}{2}+6 \arcsin^2\frac{x}{2} \bigg]
     \nonumber\\& \qquad
     +12\bigg[\mathrm{Li}_2(x^2)-2\mathrm{Re Li}_2\,
      \left( \left( x-\frac{1}{x} \right)\left( \frac{x}{2}
      -i\sqrt{1-\frac{x^2}{4}} \right) \right)
      \bigg]
      \nonumber\\ & \qquad
      +\left[ 3\pi\sqrt{4-x^2}(x^2-2)x-3(x^4+8x^2-9)x^2-5\pi^2 \right]\bigg\}
     \nonumber\\&
     -\frac{|C_{u\varphi}|^2}{\Lambda^4}\frac{(1-x^2)^2}{9\pi}\bigg[
	\left( 36\log\frac{m_t}{\mu}+4\pi^2-51 \right)+24\mathrm{Li}_2x^2
	+24\log x\log(1-x^2)
	\nonumber\\&\qquad
	+24\frac{x^2}{1-x^2}\log x+6\left( 5-\frac{2}{x^2} \right)\log(1-x^2)
	\bigg]\,.
  \label{}
\end{flalign}
\end{widetext}
The $|C_{uG}|^2$ term does not have a $\log\frac{m_t}{\mu}$ dependence.  This
is because the tree-level amplitude does not have a contribution from $O_{uG}$.
As a result, the $|C_{uG}|^2$ term entirely comes from real corrections
(virtual corrections are interferences between tree- and one-loop level
amplitudes), and it is independent of $\mu$.

In addition, the $C_{uG}^2$ term contains $\log x$ which is divergent in the
limit $m_h\to 0$. This corresponds to a soft Higgs emission which in the $m_h
\to 0 $ limit is divergent when $E_h=0$.  In this limit, we have
\begin{flalign}
  \alpha_s\Gamma^{(1)}=-\frac{\alpha_sG_F}{144\sqrt{2}\pi^2}
  \frac{|C_{uG}|^2}{\Lambda^4}m_t^7(96\log x+271)+\mathcal{O}(x)\,.
  \label{eq:logdiv}
\end{flalign}
As this term is purely from the real corrections, it can be thought of as the
real Higgs emission correction to the decay mode $t\to u+g$. The soft
divergence is expected to be canceled by the wave function renormalization of
the top quark in the process $t\to u+g$, coming from a virtual Higgs bubble
diagram.  As a check, we have computed this diagram and find
\begin{flalign}
  \delta Z_{2,h}^{(t)}=-\frac{m_t^2G_F}{16\sqrt{2}\pi^2}D_\epsilon
  \left( \frac{1}{\epsilon_{UV}}-8\log x-3 \right)+\mathcal{O}(x)\,.
\end{flalign}
The corresponding contribution to the virtual correction to the decay width of $t\to u+g$ is
\begin{flalign}
  \Gamma^{({\rm virt})}_{t\to u+g}=&
  \Gamma^{(0)}_{t\to u+g}\times \delta Z_{2,h}^{(t)}
\nonumber\\
=&\left( \frac{4\alpha_s}{3}\frac{|C_{uG}|^2}{\Lambda^4}m_t^5 \right)
\times\delta Z_{2,h}^{(t)}
\nonumber\\
=&-\frac{\alpha_sG_F}{12\sqrt{2}\pi^2}m_t^7D_\epsilon
  \left( \frac{1}{\epsilon_{UV}}-8\log x-3 \right)+\mathcal{O}(x)\,,
\end{flalign}
which exactly cancels the $\log x$ term in Eq.~(\ref{eq:logdiv}).

The other two terms in $\Gamma^{(1)}$,  $|C_{u\varphi}^2|$ and
Re($C_{uG}C_{u\varphi}^*$)  do not have this divergence. In particular, the
interference term [which is proportional to Re($C_{uG}C_{u\varphi}^*$)] is
finite in the $x\to 0$ limit, even though it contains $\log x$ terms and
$\mathrm{Li}_2$ functions. This is because the two real correction diagrams from
$O_{u\varphi}$ cancel each other when $p_h=0$.

\section{Numerical analysis}

For the numerical analysis we assume $\Lambda=1$ TeV. For the input parameters,
we use
\cite{Beringer:1900zz}
\begin{flalign}
  m_t & =173.5 \ \mathrm{GeV}\\
  m_h & =125.3 \ \mathrm{GeV}\\
  G_F & =1.1664\times10^{-5} \ \mathrm{GeV}^{-2} \,.
\end{flalign}
With these parameters we find
\begin{flalign}
  \Gamma^{(0)}=&7.11 |C_{u\varphi}(\mu)|^2\  \times  10^{-4}\, \mathrm{GeV}\,,\\
  \Gamma^{(1)}=&
  \bigg\{
  \left[1.19-9.05\log\left(\frac{m_t}{\mu}\right)\right]
    |C_{u\varphi}(\mu)|^2\nonumber\\&
    -\left[3.26+18.1\log\left(\frac{m_t}{\mu}\right)\right]
    \mathrm{Re}C_{uG}(\mu)C_{u\varphi}^*(\mu)\nonumber\\
    &+9.33\times10^{-5}|C_{uG}(\mu)|^2\
  \bigg\}
  \times 10^{-4}\, \mathrm{GeV}\,.
  \label{}
\end{flalign}
The $C_{uG}^2$ term is 4 orders of magnitude smaller than the other two
terms, and thus it is interesting to understand such a suppression. As we
have mentioned, this term only receives contributions from real emission. We
find that, due to the $\sigma^{\mu\nu}p_{g\nu}$ structure of the coupling
from $O_{uG}$, the squared amplitude for $t\to u+h+g$ depends on $p_g\cdot
p_u$. We find
\begin{flalign}
  |M|^2=&128\pi^2\alpha\alpha_s\frac{|C_{uG}|^2}{\Lambda^4}\frac{m_t^6}{m_W^2s_W^2}
  \frac{\hat{t}^2}{(1-\hat{t})^2}
  \nonumber\\
  &\times \left( \hat{t}^2+\hat{t}\hat{s}+(2-x^2)\hat{t}-\hat{s}+1 \right)
\end{flalign}
where $\hat{s}=(p_t-p_u)^2/m_t^2$ and $\hat{t}=(p_g\cdot p_u)/m_t^2$.

As a result, this term is dominated by the phase space region where $\hat{t}$
is large. However, the maximum value of $\hat{t}$ is $(1-x)^2/2$ and is
therefore suppressed for a large Higgs mass. In fact, for $m_h=125$ GeV this
suppression factor for $|M|^2$ already reaches the $10^{-3}$ level. The phase
space itself accounts for one additional order of magnitude, so the total decay
width from $|O_{uG}|^2$ is small for $m_h=125$ GeV. On the other hand, the
other two terms [$|C_{u\varphi}^2|$ and Re($C_{uG}C_{u\varphi}^*$)]  are not
affected by this factor as their main contribution comes from virtual $1\to 2$
topologies.

We now consider the impact of the NLO corrections to phenomenological applications.
In the following we assume both $C_{u\varphi}$ and $C_{uG}$ to be real. At
order $\alpha_s$ the contribution from $C_{uG}$ is even more important than
that from $C_{u\varphi}$. Neglecting the $|C_{uG}|^2$ term, the ratio between
NLO and LO result is
\begin{equation}
  \frac{\alpha_s\Gamma^{(1)}}{\Gamma_{(0)}}=0.018-0.049\frac{C_{uG}}{C_{u\varphi}}
  \label{}
\end{equation}
at $\mu=m_t$. Here we have used $\alpha_s(m_t)=0.1079$, which we obtain with
the program RunDec \cite{Chetyrkin:2000yt} from the value
$\alpha_s(m_Z)$=0.1184 \cite{Beringer:1900zz}. Without $O_{uG}$ the QCD
correction is about $2\%$, while if $O_{uG}$ and $O_{u\varphi}$ are similar in
size, the QCD correction can reach the $10\%$ level.

\begin{figure}[t]
  \begin{center}
    \includegraphics[scale=0.45]{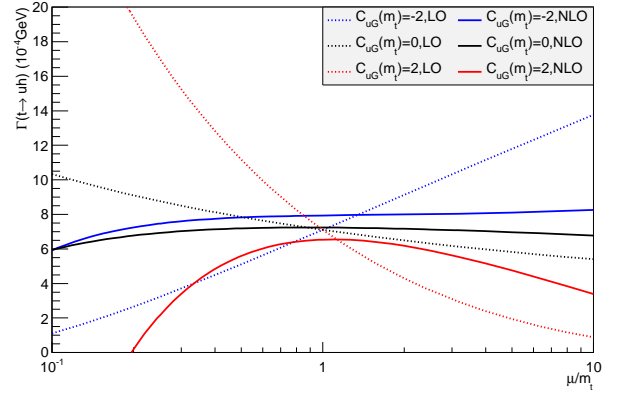}
  \end{center}
  \caption{Renormalization scale dependence of the width $\Gamma(t\to uh)$, assuming
  $C_{u\varphi}(m_t)=1$ and $\Lambda=1$ TeV.}
  \label{plot:5}
\end{figure}

\begin{figure*}[t]
  \begin{minipage}[t]{0.45\linewidth}
    \begin{center}
      \includegraphics[scale=0.45]{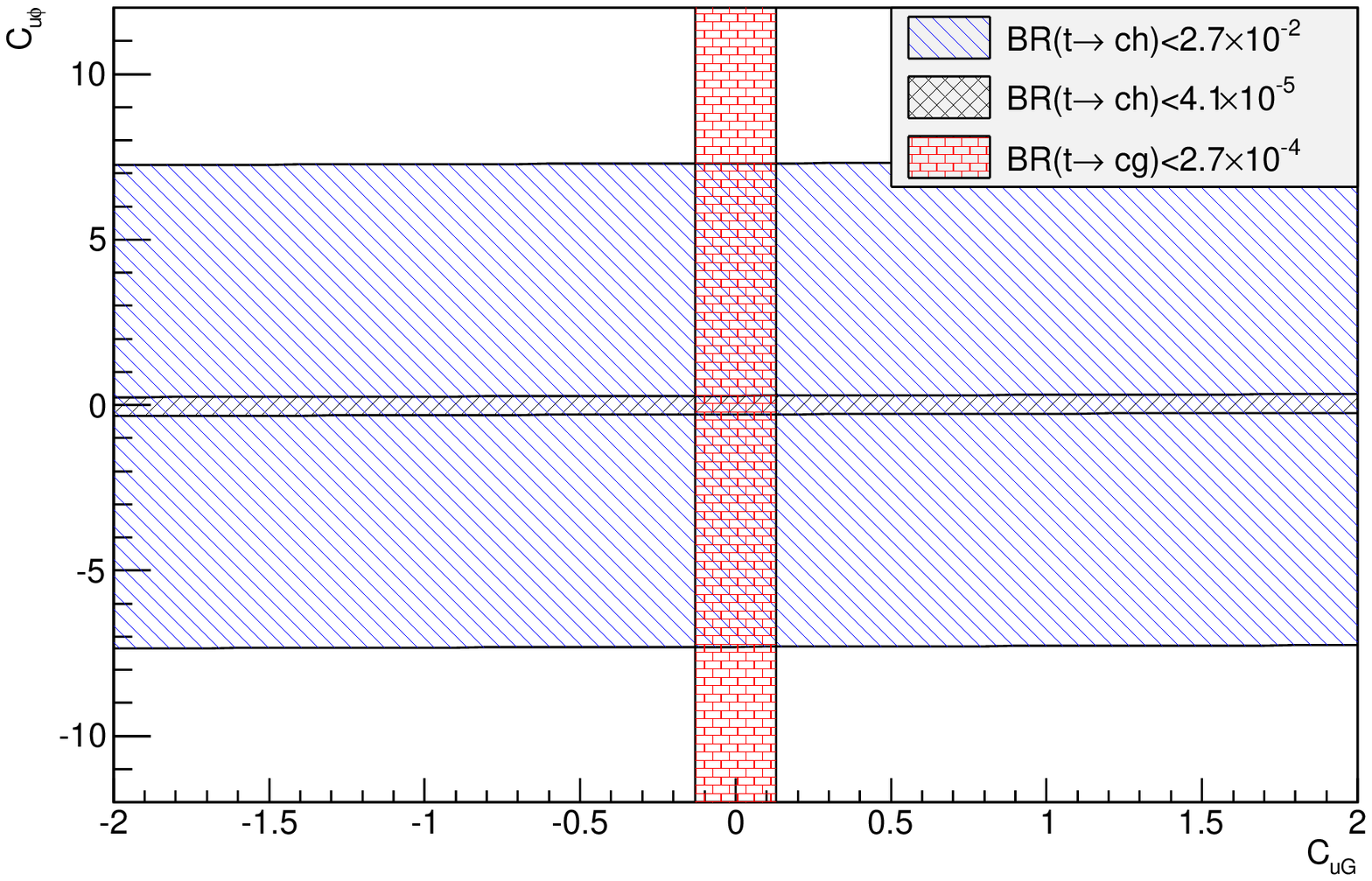}
    \end{center}
  \end{minipage}
  \hfill
  \begin{minipage}[t]{0.45\linewidth}
    \begin{center}
      \includegraphics[scale=0.45]{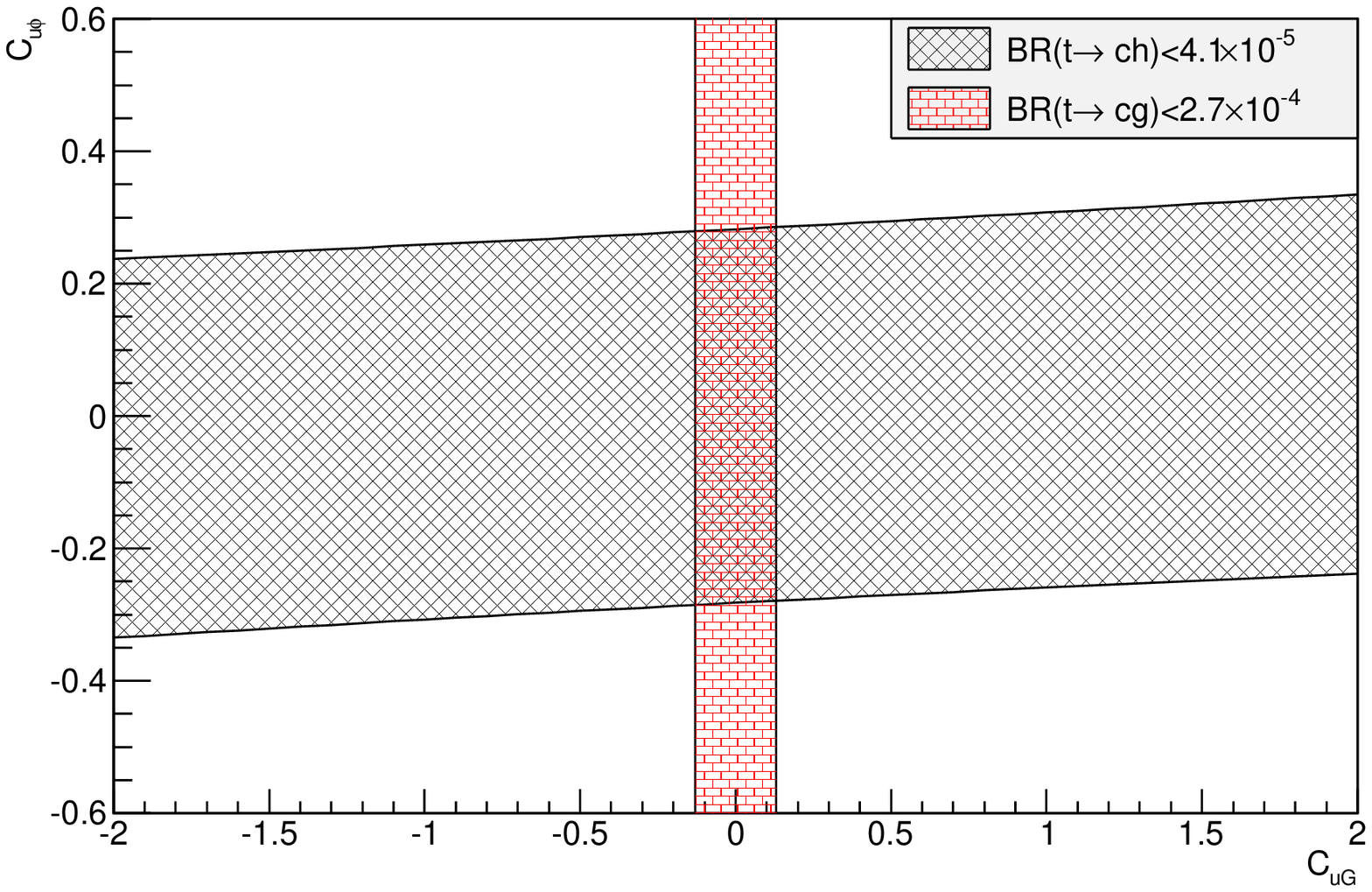}
    \end{center}
  \end{minipage}
     \caption{Limits on $C_{uG}$ and $C_{u\varphi}$ plane. Left: the blue region
   corresponds to current bound on BR$(t\to ch)$ from the LHC, the black region
 is the projected sensitivity for $t\to ch$, and the red region comes from bounds
 on $t\to cg$. Right: y-axis is zoomed in to show the effects of $C_{uG}$.\label{plot:6}}
\end{figure*}

The residual theoretical uncertainties can be estimated by checking the scale dependence of the decay width. 
Using the anomalous dimension matrix given in Eq.~(\ref{eq:AD2}), we solve the scale dependence of
the coefficients $C_{u\varphi}$ and $C_{uG}$:
\begin{flalign}
  C_{u\varphi}(\mu)=&
  C_{u\varphi}(m_t)\left( \frac{\alpha_s(\mu)}{\alpha_s(m_t)} \right) ^{\frac{4}{\beta_0}}
  \nonumber\\&
  +\frac{12}{7}C_{uG}(m_t)\left[
  \left( \frac{\alpha_s(\mu)}{\alpha_s(m_t)} \right) ^{\frac{4}{\beta_0}}
  -\left( \frac{\alpha_s(\mu)}{\alpha_s(m_t)} \right) ^{-\frac{2}{3\beta_0}}
    \right]
    \\
    C_{uG}(\mu)=&
    C_{uG}(m_t)\left( \frac{\alpha_s(\mu)}{\alpha_s(m_t)} \right) ^{-\frac{2}{3\beta_0}}\,,
  \label{}
\end{flalign}
where $\beta_0=11-2N_f/3$. The running of $C_{u\varphi}$ is affected by the
operator $O_{uG}$. In Fig.~\ref{plot:5}, we show the $\mu$ dependence of both
LO and NLO results for the width, with different values of $C_{uG}$. We can see
that the renormalization scale dependence at LO can be quite large depending on
the value of $C_{uG}$, and that it is greatly reduced at NLO in QCD.

Finally, for the sake of illustration, in Fig.~\ref{plot:6} we plot the limits
on $C_{uG}$ and $C_{u\varphi}$ plane. The region in the parameter space
corresponding to the current bound BR$(t\to ch)<2.7\%$ from CMS
\cite{Craig:2012vj} is shown, as well as the $95\%$ upper limit for $t\to ch$
as estimated in Ref.~\cite{AguilarSaavedra:2004wm}, i.e. BR$(t\to qh)<4.1\times
10^{-5}$ for an integrated luminosity of 100 $\mathrm{fb}^{-1}$. In our
results, the $|C_{uG}|^2$ term has been neglected and the
next-to-next-to-leading-order top quark width result $\Gamma(t\to bW)=1.39\
\mathrm{GeV}$ \cite{Czarnecki:2010gb} is used.  The constraints on $C_{uG}$
coming from BR$(t\to cg)<2.7\times10^{-4}$ \cite{Cristinziani:2013ila} are also
shown.

\section{Conclusion}

We have presented a calculation for the decay width of $t\to u_ih$ in the EFT
approach at NLO in QCD. Two operators contribute at LO, while at NLO two
additional operators (and their mixing) need to be included.  We find that QCD
correction can reach the 10\% level, depending on the relative size of these
operators. The possibly large scale dependence of the LO results is tamed at
NLO.

\section{Acknowledgements}
This work  is supported by the IISN ``Fundamental interactions'' convention
4.4517.08.  and in part by the Belgian Federal Science Policy Office through
the Interuniversity Attraction Pole P7/37.

\bibliography{bib}
\end{document}